\documentclass[preprint,aps,showpacs]{revtex4}
\usepackage{dcolumn}
\usepackage{bm}

\begin{document}
\preprint{}

\voffset 0.5in

\title{Calculating Quantum Transports Using Periodic Boundary Conditions}

\author{Lin-Wang Wang}
\affiliation{Computational Research Division,
Lawrence Berkeley National Laboratory, Berkeley, CA 94720
}


\date{\today}


\begin{abstract}
An efficient new method is presented to calculate the quantum transports using 
periodic boundary conditions. This new method is based on a method we developed previously, but with 
an essential change in solving the Schrodinger's equation. As a result of this change, 
the scattering states can be solved at any given energy. Compared to the previous
method, the current method is faster and numerically more stable.  The total computational 
time of the current method is similar to a conventional ground state calculation. 
Details of the procedure is presented in the current paper. 
\end{abstract}


\maketitle

\section {Introduction}

Molecular elastic quantum transport has been studied intensely in recent years
both in theory and experiment\cite{Review,transmatrix1,transmatrix2,transmatrix3,DiVentra,Reed,Chiu}.  Theoretically, the total current through a 
molecule connected by two electrodes can be calculated as:

\begin{equation}
I= {2e\over h} \int_{\mu_L}^{\mu_R} \sum_n T_n(E) dE,
\end{equation}
where $\mu_L$ and $\mu_R$ are left and right electrode Fermi energies
(assuming the current flows from right to left in z direction), 
and  $T_n(E)$ is the transmission coefficient for the $n$th right hand electrode band 
at energy E. One common way to calculate $T_n(E)$ is to calculate the scattering states
$\psi_{sc}(r)$ at energy $E$ which satisfies the Schrodinger's equation:

\begin{equation}
H\psi_{sc}(r)=E\psi_{sc}(r)  
\end{equation}
while having the following boundary conditions:

\begin{equation}
\psi_{sc}(r)=\left\{ \begin{array} {ll}
     \phi_m^{R*}(r)+ \sum_{n\ne m} B_n^{R} \phi_n^R(r)      & \mbox{ if $z\rightarrow \infty$ } \\
   \sum_n A_n^{L} \phi_n^{L*}(r)  & \mbox { if $z\rightarrow -\infty$ }
                      \end{array}
              \right. 
\end{equation}

Note that, in Eq(2), $H=\{-{1\over2}\nabla^2+V(r)+V_{nonloc}\}$ is the single particle Hamiltonian. 
In Eq(3),  
 $\phi_n^{R(L)}(r)=u_{n,k_n}(r) exp(i k_n^{R(L)}z)$, are the right going running waves 
in the the right(R) and left(L) electrodes, and $\phi_n^{R(L)*}$ are the left going 
running waves.    
$E_n^{R(L)}(k_n^{R(L)})=E$ are the electrode band structure. The summation $\sum_n$ in Eq(3) stands
for all band $n$ and $k_n^{R(L)}$ which satisfy $E_n^{R(L)}(k_n^{R(L)})=E$.  In Eq(3), we have assumed that 
${dE_n^{R(L)}(k)/dk}>0$ for band $n$. If one band $n$ has ${dE_n^{R(L)}(k)/dk}<0$, 
then its corresponding $\phi_n^{R(L)}(r)$ in Eq(3) should be replaced by $\phi_n^{R(L)*}(r)$ for that particular
band.  
Eq(3) describes an incoming running wave $\phi_m^{R*}(r)$ from the right electrode band $m$ which 
is scattered back through outgoing running waves $ B_n^{R}\phi_n^R(r)$ at the right electrode, and
transmitted into the left going running waves $A_n^{L} \phi_n^{L*}(r)$ at the left electrode.  
As a result of this scattering process, the transmission coefficient for channel m can be calculated as

\begin{equation}
T_m(E)=[\sum_n |A_n^L|^2 (dE_n^L(k)/dk)|_{k=k_n^L}]/(dE_m^R(k)/dk)|_{k=k_m^R}.
\end{equation}

Note that, due to current conservation, 
we have:

\begin{equation}
\sum_n |A_n^L|^2 (dE_n^L(k)/dk)|_{k=k_n^L}+\sum_{n\ne m} |B_n^R|^2 (dE_n^R(k)/dk)|_{k=k_n^R}=(dE_m^R(k)/dk)|_{k=k_m^R}.
\end{equation}

Normally, Eqs(2),(3) are solved by using 
transfer matrix method \cite{transmatrix1,transmatrix2} or
Lippmann-Schwinger equation \cite{DiVentra}. However, transfer matrix method could
be unstable in a multi-channel electrode \cite{transmatrix3} and it is difficult to 
deal with the nonlocal pseudopotential\cite{transmatrix2}, and the application of 
Lippmann-Schwinger equation is computationally expensive\cite{DiVentra}. 

In a previous publication \cite{prev}, we have described a new method to calculate Eqs(2),(3)
which uses a supercell with  periodic
boundary conditions, just like in a conventional ground state total 
energy calculation. In that method, a supercell eigenstates are solved using 
conventional conjugate gradient methods \cite{CG}. Then perturbations at one end of 
the electrode are introduced, and the eigenstates are recalculated using the same 
conjugate gradient method \cite{CG}. Next, these eigenstates are linearly 
recombined to make it satisfy the boundary condition Eq.(3). Although highly efficient compared to 
methods before it, and simple to implement since it uses only conventional ground state codes, 
that method has a drawback. In that method, the energy $E$ in Eq(2) can only 
be the eigen energies of the original supercell Hamiltonian H. As a result, $T_n(E)$ is only known
for a finite number of $E$ (or say $k_n$). To overcome this problem, 
one needs to fit $T_n(k_n)$ with a continuous function before it is used to 
calculate the total current in Eq(1). Although it has been shown in Ref. \onlinecite{prev}
that this fitting over the $k_n$ points is equivalent to the k-point summation 
in a supercell ground state calculation and it works fine in the case considered, but
there might be cases where denser energy $E$ points are needed, for example, close
to a weakly coupled resonant tunneling. In this paper, we will provide an essential 
modification over our previous method.  Under this new method, 
Eqs(2),(3) can be solved for arbitrary $E$, and the overall computation is faster than the
previous method for much denser $E$ point grid.
We also provide details of the whole procedure in the current paper.

\section{The formalism}

In order to compare with our previous method in Ref. \onlinecite{prev}, we choose the 
same system as studied in that paper. The system is schematically shown in Fig.1 taken from 
Ref. \onlinecite{prev}. In the system, a benzene molecule is connected by two Cu quantum wires through 
the bonds of two sulfur atoms. In Fig.1, left hand side B and right hand side B are periodically 
connected.  As in Ref. \onlinecite{prev}, we will do nonselfconsistent
calculations for finite bias, 
although selfconsistent calculation is straight forward using the scattering state solutions of Eqs(2),(3). 
To get the potential of a finite V bias system, 
self-consistent local density approximation (LDA) ground state calculation is performed for 
the zero bias system. Then, an additional smooth function is added to the potential 
to raise the right hand electrode by $V/2$ and lower the left hand electrode by $-V/2$. 
Norm conserving pseudopotentials are used, so is a planewave basis set with a 30 Ryd cutoff. 

With some minor changes of notations, the essential idea of Ref. \onlinecite{prev} [its Eq(6)] can be recast as to solve
the wavefunction $\psi_{(l)}(r)$ :

\begin{equation}
   (H-E)\psi_{(l)}(r)=W_{(l)}(r),
\end{equation}
here $W_{(l)}(r)$ are some perturbation functions which are only nonzero away from the molecule as 
shown in Fig.1. 
Note that,  $\psi_{(l)}(r)$ has a supercell k-point ${\bf K}_z$, thus $u_{(l)}(r)=\psi_{(l)}(r) exp(-i {\bf K}_z z)$ is
periodic.  ${\bf K}_z$ could be, for example $\pi/2 L_z$
where $L_z$ is the supercell length in the z direction.  
After a few $\psi_{(l)}(r)$ are solved for a same energy $E$, these $\psi_{(l)}(r)$
can be linearly recombined with proper coefficients to generate a scattering state $\psi_{sc}(r)$ which 
satisfies the boundary conditions of Eq(3). In Ref.  \onlinecite{prev}, Eq(6) is obtained by combining 
two eigen state equations with two different $W_{m}$ [ Eq(6) in Ref. \onlinecite{prev}]. Its advantage 
is that it needs only  conventional eigen state calculations, thus there is no need to 
change a ground state code. 
The disadvantage, however, is that the energy $E$ in Eq(6) can only be the eigen state energy $E_i$ of
the unperturbed ($W_{(l)}(r)=0$) supercell system. Here, we will solve the Eq(6) directly using the 
conjugate gradient method. The approach is very similar to the method used in perturbation linear 
response theory \cite{Gonze}. 

Notice that, the linear equation (6) can be rewritten as an optimization of the following  F:

\begin{equation}
 F= <\psi_{(l)}| H-E | \psi_{(l)}> - <\psi_{(l)}|W_{(l)}>-<W_{(l)}|\psi_{(l)}>.
\end{equation} 

Preconditioned conjugate gradient method can be used to solve the minimum 
of F. Unfortunately, for an arbitrary $E$, the matrix $H-E$ in the above equation 
is not positive definite, which makes the conjugate gradient method diverges. However, 
this problem can be circumvented if we have the eigenstates $\{ \psi_i, E_i \}$ of the
unperturbed Hamiltonian $H$. First, if we know all the eigenstates $\{ \psi_i, E_i \}$ of H, 
the $\psi_{(l)}$ in Eq(6) can be solved directly as: 

\begin{equation}
\psi_{(l)}=\sum_i^{\infty} {<\psi_i| W_{(l)}> \over E_i-E} \psi_i
\end{equation}

In practice, however, we usually only solve the eigenstates
$\{ \psi_i, E_i \}$ (using the conjugate gradient method \cite{CG})
up to an energy $E'$ (with $E' > \mu_R$). 
Let's denote this converged set of eigenstates as $i=1,N$. Then the 
idea is to deflate these N eigenstates from both $\psi_{(l)}$ and $W_{(l)}$, and solve the 
remaining part of the wavefunction. More specifically, we can define:

\begin{equation}
\psi^P_{(l)}=\hat P \psi_{(l)}= \psi_{(l)}-\sum_{i=1}^N <\psi_i | \psi_{(l)}> \psi_i
\end{equation}

and

\begin{equation}
W^P_{(l)}=\hat P W_{(l)}= W_{(l)}-\sum_{i=1}^N <\psi_i | W_{(l)}> \psi_i. 
\end{equation}

Then the linear equation 
\begin{equation}
   (H-E)\psi^P_{(l)}(r)=W^P_{(l)}(r),
\end{equation}
in the subspace of projector  $\hat P$ can be solved as the minimum of

\begin{equation}
 F^P= <\psi^P_{(l)}| H-E | \psi^P_{(l)}> - <\psi^P_{(l)}|W^P_{(l)}>-<W^P_{(l)}|\psi^P_{(l)}>.
\end{equation} 

Note that, now the effective matrix $\hat P (H-E) \hat P$ is positive definite as long as E is 
lower than $E_N$. When using conjugate gradient method to solve Eq(12), the projector $\hat P$ 
are repeatedly applied to the wavefunctions and search directions, so that the whole minimization is 
done within the subspace defined by $\hat P$. After $\psi^P_{(l)}$ is solved, $\psi_{(l)}$ of 
Eq(6) can be obtained from Eqs(9) and (8) as: 

\begin{equation}
\psi_{(l)}= \psi^P_{(l)}+ \sum_{i=1}^N {<\psi_i| W_{(l)}> \over E_i-E} \psi_i .
\end{equation}

The convergence of $\psi^P_{(l)}$ under the conjugate gradient method is very fast since the effective
band gap for $\psi^P_{(l)}$ is $E_N-E$. Kinetic energy G-space diagonal preconditioning can be 
used, just as in the conventional ground state conjugate gradient method\cite{CG}. Fig.2 shows a typical 
convergence for a $\psi^P_{(l)}$ state. It shows that 20 conjugate gradient line minimizations 
is enough to converge the $\psi^P_{(l)}$ to $10^{-6}$ (a.u.) starting from zero.  The $\psi^P_{(l)}$
can be converged to the same accuracy as that of $\{ \psi_i,E_i\}$. 
We find similar convergence for all $W_{(l)}$ and $E$.  

To linearly combine $\psi_{(l)}$ to generate 
$\psi_{sc}$ of Eq(3), we want to have linearly independent $\psi_{(l)}$. 
According to Eq(8), if $M$ $W_{(l)}$ are linearly independent, then the corresponding 
$M$ $\psi_{(l)}$ are also linearly independent. 
As in Ref. \onlinecite{prev}, we intend to use the $\Gamma$ point electrode states as  $W_{(l)}$. 
However, if one or few $E_i$ are very close to E, then their corresponding  $\psi_i$ terms
might dominate the expression in Eq(8), which can make the $\psi_{(l)}$ lie in similar directions
for different $l$. To avoid this situation, we have modified $W_{(l)}$ as following. Let's use $W^0_{(l)}$ to 
denote the $\Gamma$ point electrode states in real space and here $l$ is an index for different bands. 
$W^0_{(l)}$ are 
only nonzero at the last primary cell of the right electrode as shown in Fig.1. Then from $W^0_{(l)}$, 
we can generate $W_{(l)}$ using the following iterations from $\mu=1$ to $\mu=l-1$:

\begin{equation}
W_{(l),\mu+1}=W_{(l),\mu}- W_{(\mu)} {<\psi_{m_{\mu}}|W_{(l),\mu}>\over <\psi_{m_{\mu}}|W_{(\mu)}> }
\end{equation}

and $W_{(l),1}=W^0_{(l)}$ and $W_{(l)}=W_{(l),l}$. 
In the above equation, $m_{\mu}$ is the i which gives the maximum $|<\psi_i|W_{(\mu)}>/(E_i-E)|$ for a given $\mu$. Note that, it is easy to show from 
Eq(14) that $<\psi_{m_{\mu}}|W_{(l)}>=0$ for all the $\mu < l$. In other words, $\psi_{(l)}$ 
as described in Eq(8) [or Eq(13)] will not have the $\psi_i$ component if $\psi_i$ is a maximum
component in one $\psi_{(\mu)}$ with $\mu < l$. This makes $\psi_{(l)}$ and $\psi_{(\mu)}$  ($\mu < l$) 
unlikely to lie in very close directions. Note that $W_{(l)}$ from Eq(14) will still only be nonzero
in the last primary cell of the right electrode.

After the $M$ $\psi_{(l)}$ of Eq(6) are calculated following the above procedure for a given $E$, we will combine
these $\psi_{(l)}$ to generate the scattering states $\psi_{sc}$ of Eqs(2),(3). This part is similar 
to what we have done in Ref. \onlinecite{prev}. However, more details will be provided here. A band 
structure alignment between the right and left electrodes is illustrated in Fig.3 with an 4 V bias. 
The numbers in the right electrode band structure are the index of the electrode bands. In our calculation,
we have over cautiously used 11 bands as $W_{(l)}$ in Eq(6). As a result, for a given energy $E$, 
we will have 
11 $\psi_{(l)}$. Since $\psi^*_{(l)}$ (which has a $-{\bf K}_z$ instead of ${\bf K}_z$) also satisfy 
Eq(6), we end up having 22 wavefunctions to be used in the linear combination to generate $\psi_{sc}$
[in the following, we will denote all these 22 wavefunctions as $\psi_{(l)}$, with $l=1,M$ and $M$ 
being 22].  
As described in Ref. \onlinecite{prev}, we will first decompose each $\psi_{(l)}$ at left and right electrode
primary cells $\Omega_L$ and $\Omega_R$ by the electrode wavefunctions. As shown in Fig.3, for a given 
energy $E$, we can find the corresponding $k_n^L(E)$ and $k_n^R(E)$ (the small black dots in Fig.3 
on the dashed line E).  If the numbers of $k$ in the left
and right electrodes are $N_L$ and $N_R$ respectively, 
then there will be $2(N_L+N_R)$ electrode running waves states (counting
also the $-k$). Like in Eq(3), let's use $\phi_n^{R(L)}$ and $\psi_n^{R(L)*}$ to denote these 
electrode running wave states, then the decomposition of $\psi_{(l)}$ can be written as:

\begin{equation}
\psi_{(l)}(r)=\left\{ \begin{array} {ll}
     \sum_{n=1}^{N_R}[A_n^R(l) \phi_n^{R*}(r)+ B_n^{R}(l) \phi_n^R(r)]      & \mbox{ if $z\in \Omega_R$ } \\
     \sum_{n=1}^{N_L}[A_n^L(l) \phi_n^{L*}(r)+ B_n^{L}(l) \phi_n^L(r)]      & \mbox{ if $z\in \Omega_L$ } 
       \end{array}
              \right. 
\end{equation}   

The coefficients $A_n^R$, $B_n^R$ can be calculated by the overlap matrix $<\phi_n^{R(*)}|\phi_m^{R(*)}>$
and the projection matrix $<\psi_{(l)}|\phi_n^{R(*)}>$. The same for $A_n^L$, $B_n^L$. The electrode
wavefunctions $\phi_n(r)$ are pre-calculated at 50 $k$ points. Then the $\phi_n^R(r)$ for a given 
$k_n^R$ point is obtained vs interpolation.  

Now, combining $\psi_{(l)}(r)$ we have the scattering state as:

\begin{equation}
\psi_{sc}(r)=\sum_{l=1}^M C_l \psi_{(l)}(r). 
\end{equation}

Note, $\psi_{sc}(r)$ satisfies the Schrodinger equation (2) in the region 
where $W_{(l)}(r)$ are zero (away from B in Fig.1).  
According to Eq(15), we have $\psi_{sc}(r)$ at $\Omega_R$ and $\Omega_L$ as:

\begin{equation}
\psi_{sc}(r)=\left\{ \begin{array} {ll}
     \sum_{n=1}^{N_R}\{ [\sum_{l=1}^M A_n^R(l) C_l] \phi_n^{R*}(r)+ [\sum_{l=1}^M B_n^{R}(l) C_l] \phi_n^R(r)\}      & \mbox{ if $z\in \Omega_R$ }  \\
     \sum_{n=1}^{N_L}\{[\sum_{l=1}^M A_n^L(l) C_l] \phi_n^{L*}(r)+ [\sum_{l=1}^M B_n^{L}(l) C_l] \phi_n^L(r)\}      & \mbox{ if $z\in \Omega_L$ } 

       \end{array}
              \right. 
\end{equation}   

Comparing this equation with the boundary equation (3), we have the following $N_R+N_L$ linear equations for 
a scattering state based on a $\phi_m^{R*}$ incoming wave:
        
\begin{equation}
\begin{array} {ll}
   \sum_{l=1}^M A_n^R(l) C_l = \delta_{n,m}  &   \mbox{ for $n=1,N_R$}  \\
   \sum_{l=1}^M B_n^L(l) C_l = 0             &   \mbox{ for $n=1,N_L$} 
\end{array}
\end{equation}

Note that if $M \ge N_R+N_L$ there can be a solution for the above equation. Given the 11 bands
we used as $W_{(l)}$ ($M=22$), we find that this is always true. When $M > N_R+N_L$, Eq(18) is 
under determined, meaning there are more than one solutions of $C_l$. In this case, it makes sense
to require the minimum of $\sum_{l=1}^M |C_l|^2$ while Eq(18) is satisfied. This linear
algebra problem can be solved using standard numerical routines, like the ZGELSS in LAPACK\cite{LAPACK}.  

After Eq(18) is solved, then we will have a scattering wave $\psi_{sc}$, which satisfies the
Schrodinger's equation (2) within the region from $\Omega_L$ to $\Omega_R$, and the boundary condition
of Eq(3) at $\Omega_L$  and $\Omega_R$. We can discard the $\psi_{sc}$ of Eq(16) for the 
regions outside $\Omega_L$ and $\Omega_R$ (near boundary B). Instead, the real scattering state
can be extended following the propagations of the nonzero electrode running waves in Eq(17) into 
negative and positive infinities. As a result, the boundary conditions of Eqs(17),(18) at $\Omega_R$ and 
$\Omega_L$ are the same as the boundary conditions of Eq(3) at $z\rightarrow \infty$ and 
$z\rightarrow -\infty$.

\section {The evanescent states}

Above discussions are complete if the electrodes are sufficiently long, so there are no evanescent states
at $\Omega_L$ and $\Omega_R$. In practices, however, we found evanescent states often exist. There
are two type of evanescent states. The first type (type I) originates from the molecule and decays out in the 
electrodes (i.e. $e^{-\kappa z}$ in the right electrode and $e^{\kappa z}$ in the left electrode). 
The second type (type II) originates from the artificial boundary B in Fig.1, and decays towards the molecule 
(i.e. $e^{\kappa z}$ in the right electrode and $e^{-\kappa z}$ in the left electrode). While the first 
type evanescent states could be physical, existing in a scattering state $\psi_{sc}$, the second type 
of evanescent states are artificial due to our use of boundary condition and perturbation $W_{(l)}$ near B . If we can calculate the running wave coefficients $A_n^{R(L)}(l)$ 
and $B_n^{R(L)}(l)$, then even if we ignore the evanescent states in Eq(15), our resulting scattering 
state $\psi_{sc}$ constructed from Eq(16) and Eq(18) will still be correct. This is because the evanescent
states can be added in as additional terms in Eq(17). As we extend our boundary condition from 
$\Omega_L$ and $\Omega_R$ to $-\infty$ and $\infty$, the first type evanescent states will decay out, and 
we can simply remove (subtract out) the second type evanescent states without affecting the Schrodinger's
Equation (2) (assuming its amplitude near the molecule is 
sufficiently small).  As a result, we will still have a boundary condition as in Eq(3). 

However, it is helpful to include the evanescent states in the decomposition Eq(15) for two reasons:
(1) To accurately calculate the running wave coefficients $A_n^{R(L)}(l)$ and $B_n^{R(L)}(l)$;
(2) In Eq(16), to avoid the case where large artificial second type evanescent
states exist and dominate the equation.  As a result, they have significant 
tail amplitudes near the molecule (compared to the running wave amplitudes). If this is true, then 
these second type evanescent states cannot be simply removed without introducing errors. 

To get $A_n^{R}(l)$ and $B_n^{R}(l)$ from Eq(15), we have used the overlap matrix 
$<\phi_n^R|\phi_m^R>_{\Omega_R}$, $<\phi_n^{R*}|\phi_m^R>_{\Omega_R}$, 
$<\phi_n^{R*}|\phi_m^{R*}>_{\Omega_R}$ and projections $<\psi_{l}|\phi_n^R>_{\Omega_R}$ 
and $<\psi_{l}|\phi_n^{R*}>_{\Omega_R}$, then solved the resulting linear equation. Here the subscript
$\Omega_R$ means the integration is done only within $\Omega_R$. Since the running waves
and evanescent states are not orthogonal within $\Omega_R$, then ignoring the evanescent states 
in Eq(15) will introduce errors in the resulting $A_n^{R}(l)$ and $B_n^{R}(l)$. The situation is
the same for the left electrode.

The evanescent states are originated from the real $k_e$ points when $dE_n(k)/dk|_{k=k_e}=0$. 
Here the subscript $e$ stands for evanescent states. This happens
at the $\Gamma$ and $X'$ points of the electrode band structure as shown in Fig.3, and at the place 
where two bands anticross each other and form a band gap. At the $\Gamma$ point, an evanescent state line
runs downward starting from a real $\Gamma$ point band structure energy.  At the $X'$ point 
and other anticrossing points, two evanescent state lines connect the two real $k_e$ points at the opposite edges of an
energy gap. A detail description of the complex band structure is given by Chang in 
Ref. \onlinecite{Chang}. Although calculating the complex band structure is possible \cite{Chang}, 
it is difficult for a nonlocal pseudopotential Hamiltonian as is used here. As a result, we have 
used the real $k_e$ point Bloch states $\phi_{n,e}(r)=u_{n,k_e} exp(i k_e z)$ 
to approximate the evanescent states. Notice that, these are just normal running wave states, except
that they carry no current since $dE_n(k)/dk|_{k=k_e}=0$. A more accurate approximation is to add an
exponential decaying factor $exp( \kappa z)$ or $exp(-\kappa z)$ to $\phi_{n,e}(r)$. However, since we are only 
going to use $\phi_{n,e}(r)$ within $\Omega_R$ or $\Omega_L$, and the surviving evanescent
states within $\Omega_R$ or $\Omega_L$ should have a small $\kappa$, we found it is okay for not using 
these decaying factors.  
By not adding this decaying factors, we also do not distinguish the type one and type two evanescent 
states.   

Unlike the running wave states whose number is finite for a given energy $E$, there can be many 
(actually infinite if we have an infinite basis set) evanescent states for a given $E$. 
This is because at the $\Gamma$ point of the band structure, every new band will have an evanescent 
state line running downward in energy \cite{Chang}. As a result, in Eq(15), we cannot include all the possible 
evanescent states $\phi_{n,e}(r)=u_{n,k_e} exp(i k_e z)$ for a given $E$. On the other hand, 
in practice, it is not necessary
to include the evanescent states which are originated from running wave energies $E_n(k_e)$ which 
are far away from $E$,   
because they will have fast decay factors $exp( \kappa z)$, thus should not exist in $\Omega_L$ or $\Omega_R$.  
Because of this, we have the following practical procedure in solving Eq(15) and selectively including 
the evanescent states(we will use the right electrode as the example, the same is true for the 
left electrode): (1) We will start with all the running wave states, calculate the overlap matrix
elements $<\phi_n^R|\phi_m^R>_{\Omega_R}$, $<\phi_n^{R*}|\phi_m^R>_{\Omega_R}$, 
$<\phi_n^{R*}|\phi_m^{R*}>_{\Omega_R}$ and projections $<\psi_{l}|\phi_n^R>_{\Omega_R}$, then solve
the linear equations for $A_n^R(l)$, $B_n^R(l)$. (2) We will calculate the integral of wavefunction square of
the right and left hand sides of Eq(15) within $\Omega_R$, and calculate the percentage of the
right hand side vs the left hand side results. We will call this decomposition percentage (which is alway
less than 1). If this percentage is close to 1 within a criterion (e.g., $10^{-4}$), then stop. Otherwise
go to next step. (3) We will include the next evanescent state which has its $E_n(k_e)$ closest to $E$.  
We will include $\phi_{n,e}(r)=u_{n,k_e} exp(i k_e z)$ in summation of Eq(15), just treat it as one 
of the running wave states (but if $k_e$ is $\Gamma$ or $X'$, $\phi_{n,e}^*(r)$ is the same as 
$\phi_{n,e}(r)$, thus should not be included). Then repeat step (1),(2), find the new $A_n^R(l)$, $B_n^R(l)$, also the values for
the evanescent states $A_{n,e}^R(l)$, $B_{n,e}^R(l)$. If one evanescent state has almost zero (e.g., less than 
$10^{-4}$) contributions in all $\psi_{(l)}(r)$, then discard this evanescent state. If the 
decomposition percentage is still not close enough to 1, repeat step (3). 
If the total number of evanescent state is too big (e.g, larger than 10), or the next closest $E_n(k_e)$
is too far away from $E$ (e.g, farther than 2 eV), then stop. 

Let's assume that through the above procedure, we have included $N^e_L$, $N^e_R$ evanescent states
(counting both possible $\phi_{n,e}$ and $\phi_{n,e}^*$) to 
the $\Omega_L$ and $\Omega_R$ sub-equations in Eq(15). If we have $M> N_R+N_L+N^e_R+N^e_L$ (situation I), then 
we can request all the evanescent state coefficients to be zero after the linear combination in Eq(17) [e.g,
$\sum_{l=1}^M A_{n,e}^{R(L)}(l)C_l=0$, $\sum_{l=1}^M B_{n,e}^{R(L)}(l)C_l=0$ ]. These are $N_L^e+N_R^e$
additional equations to Eq(18).  
Since we still have more number of $C_l$ than the total number of linear equations, 
we can still request the $\sum_l^M |C_l|^2/ \omega_l$ to be minimum while these equations are 
satisfied (again, this can be solved by the ZGELSS LAPACK routine \cite{LAPACK}).  
Here we have placed a weight function $\omega_l$, which depends on the decomposition
percentage (after the inclusion of the evanescent states) of each $\psi_{(l)}$ in Eq(15). 
If the decomposition 
percentage is close to 1 [a good fit in Eq(15)], then $\omega_l$ is close to 1. If the decomposition percentage is
much less than 1 [not a very good fit in Eq(15)], then  $\omega_l$ is very small, which means $\psi_{(l)}$ is discouraged from
participating in the linear combination of Eq(16). 
More specifically,
if we use $p_l^R$ and $p_l^L$ to denote the decomposition percentage 
of the $\psi_{(l)}$ at $\Omega_R$ and $\Omega_L$ in Eq(15), then 
we have used a formula  $\omega_l=0.001/(0.001+|p_l^R-1|)+0.001/(0.001+|p_l^L-1|)$. 
In another situation (situation II), 
we have  $N_R+N_L \le  M \le N_R+N_L+N^e_R +N^e_L$. Then to solve $C_l$, 
we can minimize the evanescent state coefficients after the linear combination of Eq(16) 
[i.e, minimize $\sum_{n,R,L} |\sum_{l=1}^M A_{n,e}^{R(L)}(l)C_l|^2+|\sum_{l=1}^M B_{n,e}^{R(L)}(l)C_l|^2$]
while satisfying Eq(18) exactly. This again can be solved by standard numerical packages.  
In our calculation, we find all of our cases fall into situation I. 

In the situation I, we have requested the evanescent state coefficients in the  scattering state
of  Eq(16) to be zero. This might look strange at first. As we discussed above, the type I evanescent
state might be physical in a scattering state. Then, how can we force it to be zero and still have 
a good scattering state? The answer lies in the fact that we did not separate the evanescence state
$\phi_{n,e}(r)=u_{n,k_e} exp(i k_e z)$ into the type I and type II states [ i.e.,  $\phi_{n,e} e^{\kappa z}$ and 
$\phi_{n,e} e^{-\kappa z}$]. As a result, the coefficient we have for $\phi_{n,e}(r)$ is really the
sum of the coefficients for the type I and type II states. As a result, in
a scattering state $\psi_{sc}$, although we cannot force the coefficients of type I evanescent states 
to be zero, we can always add a type II states to cancel their coefficients. So, when we require the 
coefficients of $\phi_{n,e}(r)$ to be zero, it doesn't  mean the type I evanescent state coefficients
are zero.  However, since the possible type I evanescent state coefficient in a given scattering 
state $\psi_{sc}$ is fixed and is likely small at $\Omega_L$ and $\Omega_R$, then our type II
evanescent state coefficient should also be small.  This guarantees that the erroneous situation of
 large type II 
evanescent states (as we discussed near the beginning of this section)
will never happen, and our results are always stable and accurate.

\section {The results}

Following the above procedures, we have calculated the system in Fig.1 with different biases. We have
compared the current results with the results reported in Ref. \onlinecite{prev}.
First, using the conventional ground
state conjugate gradient program \cite{PEtot}, we have solved all the eigen states of $H$ up to $\sim$0.5 eV 
above the right electrode Fermi energy $\mu_R$. In our system, this amounts to $\sim$140 eigen states. 
Then we have scanned the 
scattering state energy $E$ with an interval of $\sim$ 0.04 eV. For each $l$ and $E$,  as shown in Fig.2, 
Eq(11) of $\psi_{(l)}^P$
can be solved by the conjugate gradient method  within $20$ line minimizations up to $10^{-6}$ a.u. accuracy. Next, decomposition of
$\psi_{(l)}$ is carried out at $\Omega_R$ and $\Omega_L$ as described by Eq(15) including the evanescent
states. We find that, for most $\psi_{(l)}$, the running wave alone can get a decomposition 
percentage up to $99.999\%$ or higher. However, for each energy $E$,  
it is very likely that there are one or two $\psi_{(l)}$ with their running wave decomposition 
percentage only up to $50\%$ or smaller. It is also likely that, even after including the 
approximated evanescent states, there are still one or two $\psi_{(l)}$ with their decomposition percentage 
only being around $90\%$. However, since these $\psi_{(l)}$ have very small $\omega_l$ in  
the minimization of $\sum_l^M|C_l|^2/ \omega_l$, their $C_l$ are often exceedingly small (e.g, 
$<10^{-10}$) in the linear combination of Eq(15). Following the procedure described above, a scattering
state $\psi_{sc}$ of Eqs(16),(17),(18) is solved for each right electrode running wave $\phi_m^{R*}$. 
The typical $\psi_{sc}$ wavefunctions look the same as illustrated in Fig.3 of Ref. \onlinecite{prev}. 
The transmission coefficient $T_m(E)$ is calculated for the scattering state according to Eq(4). We find
that the current conservation equation (5) is mostly satisfied beyond $99.9\%$, and in many cases 
beyond $99.999\%$, an indication of the numerical accuracy of this approach. However, there are 
occasional and distinctive cases where 
Eq(5) is not satisfied at all (e.g., the sum of transmission and reflection is $10^5$, instead of 1). 
In these cases, the evanescent states in the constructed scattering state of Eq(16) are not eliminated, 
but dominating, perhaps due to our approximations in the treatment of the evanescent states. 
Fortunately, these cases are very rare and can be easily detected, thus to be discarded. 

Figure 4 shows the calculated transmission coefficients $T_n(E)$ for different band $n$, plotted 
as functions of the $k_z$ points. The system has a bias of 4V.  Each cross symbol corresponds to 
a calculated scattering state $\psi_{sc}$.  We have also plotted the calculated $T_n(E)$ using our
previous method \cite{prev} as rectangular symbols. As we can see, the current method and the 
previous method yield the same $T_n$ amplitudes. This is a cross check for the robustness of these two 
methods. However, since the $E$ in the previous method can only be the eigen energies $E_i$ of the 
original $H$, it only yields a finite number of the scattering states. In contrast, the current 
method can have as many scattering states as we want. Actually, as shown in Fig.4, there are 
cases (e.g, for $n=5$ and for the dip near $k_z=0.9 \pi/a$ of $n=1$) where the previous method might not
have enough calculated points to reveal the sillenXX nature of the $T_n(k_z)$ curve. In terms of the
computational time, we find the current method is faster than the previous method, despite the
fact that we now have much more data points. We find that for the number of $E$ points we used, 
the total time spent to solve Eq(11) for 
all the energies and $l$ is about 2 times the time spent to solve all the ground states $\psi_i$ of 
the original Hamiltonian $H$. This makes our transport computational time in the same order
as  a typical ground state calculation.

The points in Fig.4 are fitted by smooth curves $T_n(k_z)$ as described in Ref. \onlinecite{prev}, and the 
resulting curves are used to plot the total transmission coefficients $T(E)=\sum_n T_n(k_z(E))$, which is shown in Fig.5. Again, we have compared our 
current results with our  previous results \cite{prev} for the biases of 1 V and 4 V cases. We see that, overall, 
they are almost the same. But there are some differences in the detail. In the bias 1 V case, near 
$E-E_F=-1.5$ eV, the current method produce a well shape dip. This is due to a gap at the $X'$ point of the 
left electrode near $E-\mu_L=-0.5$ eV. The previous method missed this dip because it doesn't have a
data point with its energy falls into this left electrode energy gap. 
In the bias 4 V case, near $E-E_F=0$ eV, the current 
$T(E)$ is lower than the previous results. This is because in the $n=5$ band of Fig.4, the previous
method has only two points at the $k_z<0.6 \pi/a$ region. This leads to a fitted $T_n(k_z)$ which is too high 
compared to the correct result, and consequently an over estimated $T(E)$ near $E-E_F=0$. 

Despite the above differences between the current $T(E)$ and the previous results, their calculated
total currents are very similar. For example, in the cases of 1 V and 4 V biases, the current method
produces currents 0.0390 and 0.376 $ e^2 V/h$ respectively, while the previous method produces
0.0417 and 0.398 $e^2 V/h$ respectively. The differences are only about $5\%$. The I-V curve produced by 
the current method is very close to the result of the previous method, which is shown in 
Fig.6 of Ref.\onlinecite{prev}.     

\section {CONCLUSION}

In conclusion, we have presented a new approach to calculate the quantum transports. The current method
is based on a previous method \cite{prev} which uses the periodic boundary conditions, thus makes it 
possible to use the popular pseudopotentials and planewave basis set. Compared to 
the previous method \cite{prev}, however, the current method uses a different way to solve the periodic
wavefunction $\psi_{(l)}$ of Eq(6). As a result, the scattering states can be calculated at any given 
energy $E$. This provides a more robust way to calculate the scattering state wavefunctions and their
transmission coefficients. Under the current method,  the total computational time for a transport
problem is in the same order as the computational time of its corresponding ground state problem. Enough details of the 
procedure is presented which makes the implementation of this method possible.

\begin{acknowledgments}
This work was supported by U.S. Department
of Energy under Contract No. DE-AC03-76SF00098. This research
used the resources of the National Energy Research Scientific
Computing Center at Lawrence Berkeley National Laboratory. 
\end{acknowledgments}


\newpage

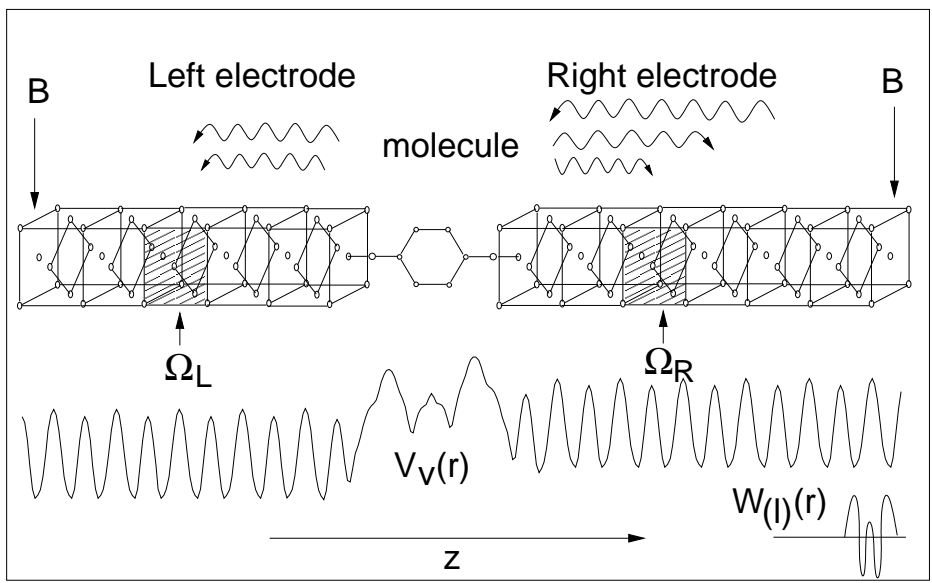
\begin{figure}[floatfix]
\caption{
A schematic view of the calculated system. }
\label{Fig1}
\end{figure}

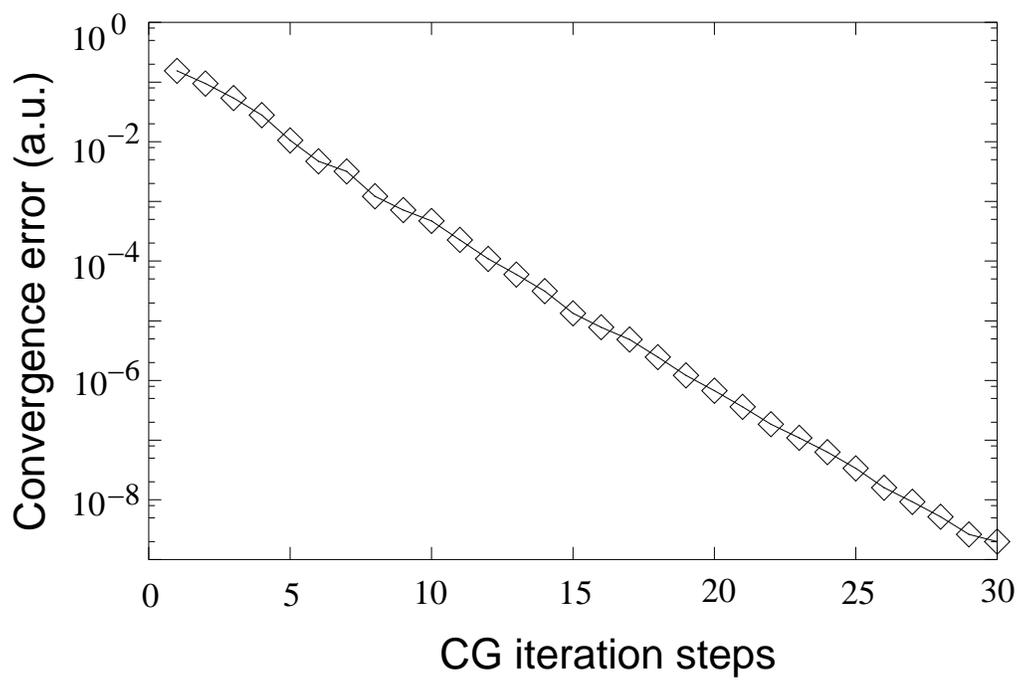
\begin{figure}[floatfix]
\caption{
The conjugate gradient (CG) convergence of Eq(11). The convergence 
error is defined as $\parallel (H-E)\psi_{(l)}^P-W_{(l)}^P \parallel$. }
\label{Fig2}
\end{figure}

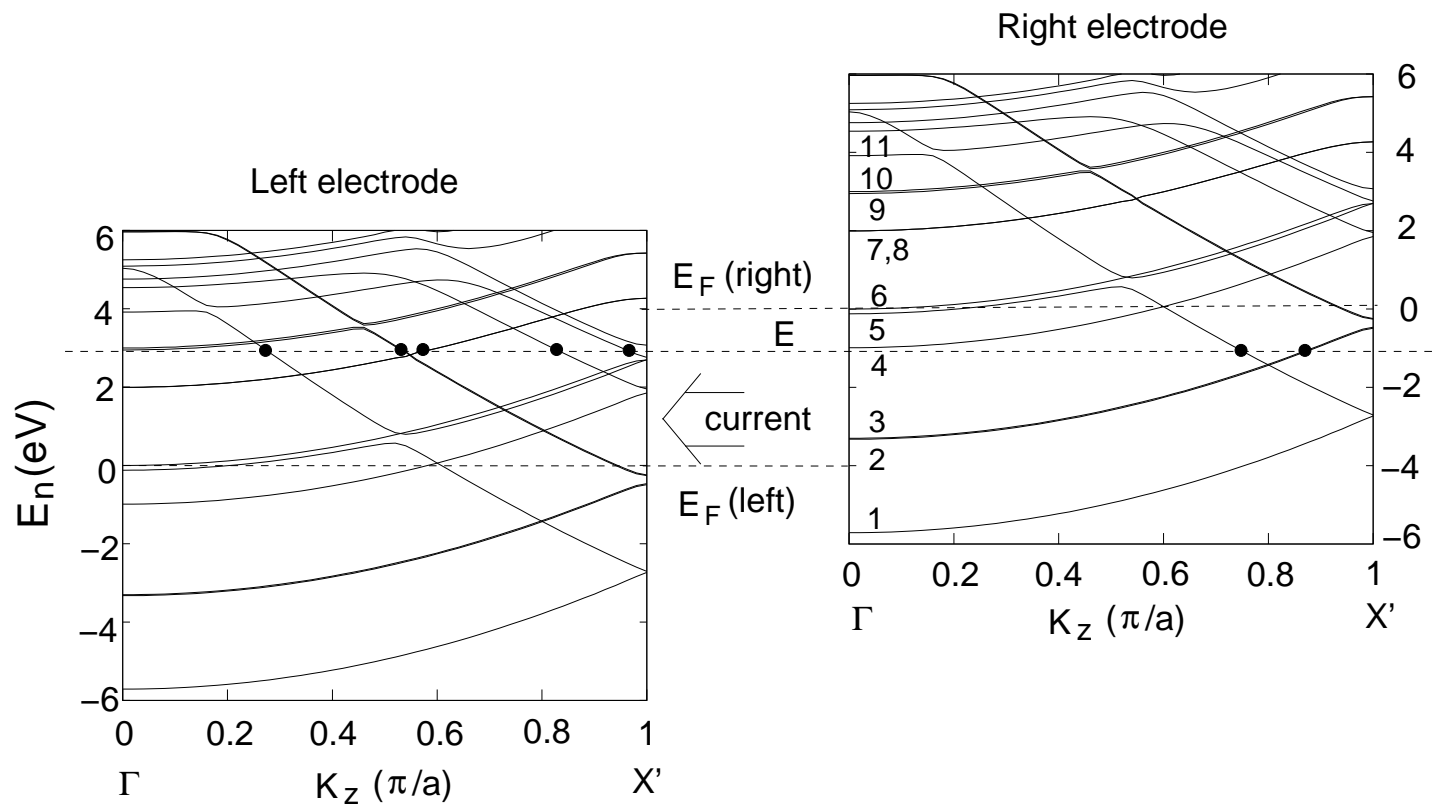
\begin{figure}[floatfix]
\caption{
The band alignment between the left electrode band structure 
and the right electrode band structure. The voltage bias is 4 V.  The numbers in the right electrode 
band structure are the band index.  The small black dots on the line E are
the $k_n^L$ and $k_n^R$ points satisfying $E_n^L(k_n^L)=E$ and 
$E_n^R(k_n^R)=E$ respectively. } 
\label{Fig3}
\end{figure}

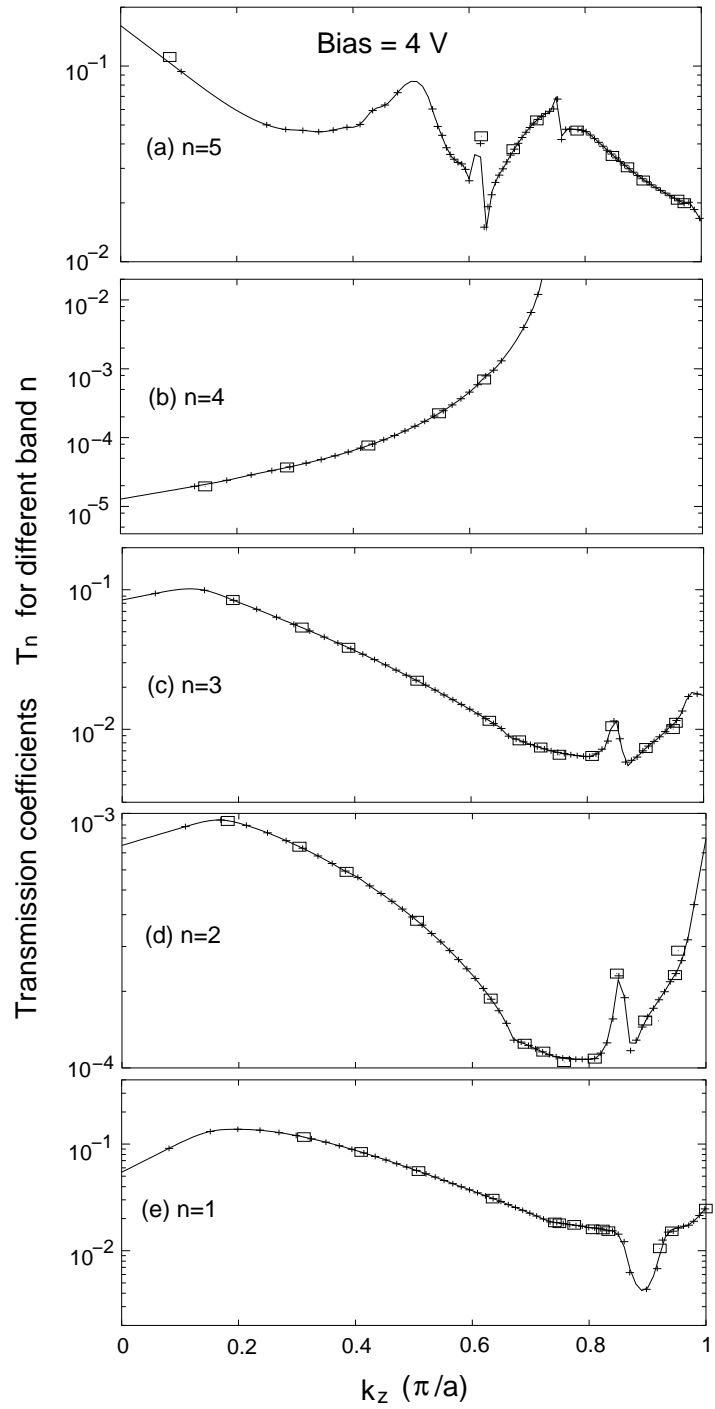
\begin{figure}[floatfix]
\caption{
The calculated transmission coefficients $T_n(k_n^R)$. The crosses are 
the results from the current method, the rectangulars are the results from 
the previous method Ref. \onlinecite{prev}, and the lines are the fitted smooth 
curves for the current results.}
\label{Fig4}
\end{figure}

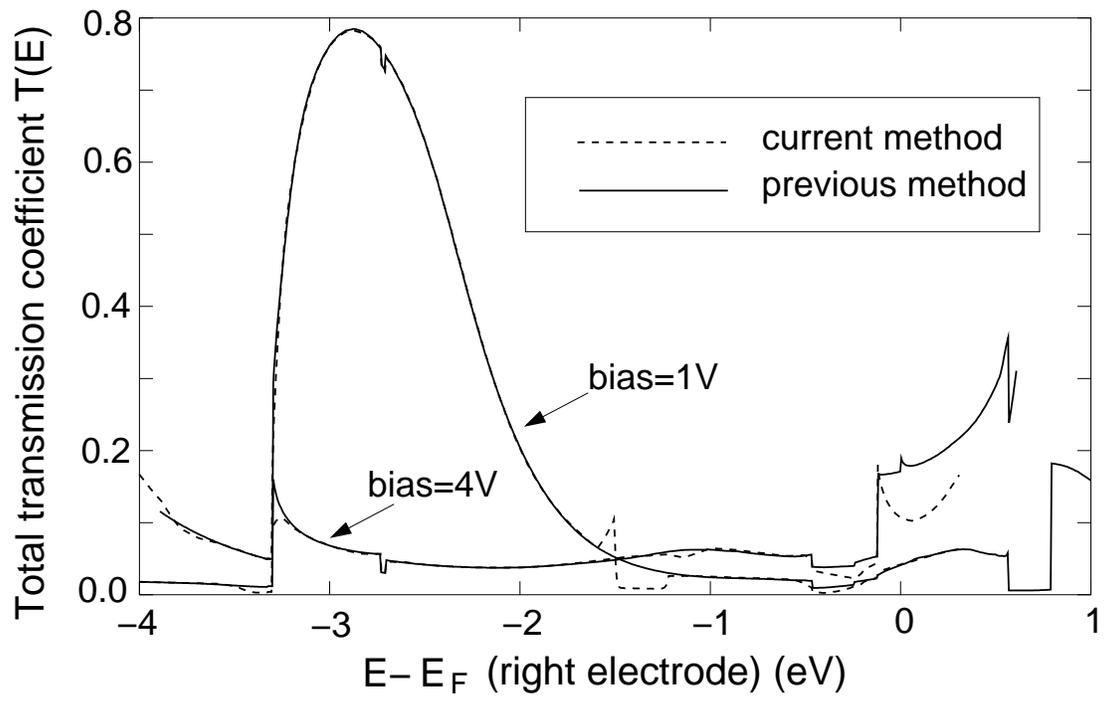
\begin{figure}[floatfix]
\caption{
The calculated total transmission coefficients $T(E)$ 
for different biases. The zero is the right electrode
Fermi energy. For a given bias V, there are net right to left current flow
only within the $[-V,0]$ energy window.}
\label{Fig5}
\end{figure}

\end{document}